\documentclass[11pt]{article}
\usepackage{times}
\usepackage{sidecap}
\usepackage{geometry}
\usepackage{floatrow}
\usepackage{float}
\usepackage{hyperref}
\usepackage{graphics,graphicx}
\usepackage{subcaption}
\usepackage{epsf,wrapfig}
\usepackage{titlesec}
\usepackage{wrapfig}
\usepackage{booktabs}
\usepackage{array}
\usepackage{amssymb}

\usepackage{natbib,bm}

\usepackage{paralist}
\let\olditem\item
\renewenvironment{thebibliography}[1]{%
  \section*{\refname}
  \let\par\relax
  \renewcommand{\item}[1][]{\olditem[\textbullet]}%
  \inparaenum}{\endinparaenum}
\usepackage{enumitem}

\newcounter{aff}

\renewenvironment{thebibliography}[1]{%
\textsc{\textbf{References:}} 
\let\par\relax%
\inparaitem[{[}1{]}]}{\endinparaitem} 

\DeclareGraphicsExtensions{.jpg,.pdf,.png,.eps}

\usepackage{caption}
\usepackage{xcolor}

\usepackage{tcolorbox}

\bibpunct{(}{)}{;}{a}{}{,}

\usepackage{soul}

\definecolor{darkcyan}{rgb}{0.0, 0.0, 0.7}

\definecolor{boxcolor}{RGB}{193, 229, 255}
\definecolor{figboxcolor}{RGB}{193, 229, 255}
\definecolor{boxcolorgray}{RGB}{227, 227, 227}
\definecolor{darkcyan}{rgb}{0.0, 0.0, 0.7}

\sethlcolor{boxcolorgray}

\def\gtrsim{\lower 2pt \hbox{$\, \buildrel {\scriptstyle >}\over
{\scriptstyle \sim}\,$}}
\def\lesssim{\lower 2pt \hbox{$\, \buildrel {\scriptstyle <}\over
{\scriptstyle \sim}\,$}}

\def\approxlt{\lower.2em\hbox{$\buildrel < \over \sim$}}
\def\approxgt{\lower.2em\hbox{$\buildrel > \over \sim$}}

\def\be{\begin{equation}}
\def\ee{\end{equation}}
\def\bea{\begin{eqnarray*}}
\def\eea{\end{eqnarray*}}

\def\jref@jnl#1{{\rm#1}}
\def\aj{\jref@jnl{AJ}}			
\def\araa{\jref@jnl{ARA\&A}}		
\def\apj{\jref@jnl{ApJ}}			
\def\apjl{\jref@jnl{ApJ}}		
\def\apjs{\jref@jnl{ApJS}}		
\def\ao{\jref@jnl{Appl.~Opt.}}		
\def\apss{\jref@jnl{Ap\&SS}}		
\def\aap{\jref@jnl{A\&A}}		
\def\aapr{\jref@jnl{A\&A~Rev.}}		
\def\aaps{\jref@jnl{A\&AS}}		
\def\azh{\jref@jnl{AZh}}			
\def\baas{\jref@jnl{BAAS}}		
\def\jrasc{\jref@jnl{JRASC}}		
\def\memras{\jref@jnl{MmRAS}}		
\def\mnras{\jref@jnl{MNRAS}}		
\def\pra{\jref@jnl{Phys.~Rev.~A}}	
\def\prb{\jref@jnl{Phys.~Rev.~B}}	
\def\prc{\jref@jnl{Phys.~Rev.~C}}	
\def\prd{\jref@jnl{Phys.~Rev.~D}}	
\def\pre{\jref@jnl{Phys.~Rev.~E}}	
\def\prl{\jref@jnl{Phys.~Rev.~Lett.}}	
\def\pasp{\jref@jnl{PASP}}		
\def\pasj{\jref@jnl{PASJ}}		
\def\qjras{\jref@jnl{QJRAS}}		
\def\skytel{\jref@jnl{S\&T}}		
\def\solphys{\jref@jnl{Sol.~Phys.}}	
\def\sovast{\jref@jnl{Soviet~Ast.}}	
\def\ssr{\jref@jnl{Space~Sci.~Rev.}}	
\def\zap{\jref@jnl{ZAp}}			
\def\nat{\jref@jnl{Nature}}		
\def\iaucirc{\jref@jnl{IAU~Circ.}}
\def\aplett{\jref@jnl{Astrophys.~Lett.}}
\def\apspr{\jref@jnl{Astrophys.~Space~Phys.~Res.}}
\def\bain{\jref@jnl{Bull.~Astron.~Inst.~Netherlands}}
\def\fcp{\jref@jnl{Fund.~Cosmic~Phys.}}
\def\gca{\jref@jnl{Geochim.~Cosmochim.~Acta}}
\def\grl{\jref@jnl{Geophys.~Res.~Lett.}}
\def\jcp{\jref@jnl{J.~Chem.~Phys.}}	
\def\jgr{\jref@jnl{J.~Geophys.~Res.}}	
\def\jqsrt{\jref@jnl{J.~Quant.~Spec.~Radiat.~Transf.}}
\def\memsai{\jref@jnl{Mem.~Soc.~Astron.~Italiana}}
\def\nphysa{\jref@jnl{Nucl.~Phys.~A}}
\def\physrep{\jref@jnl{Phys.~Rep.}}
\def\physscr{\jref@jnl{Phys.~Scr}}
\def\planss{\jref@jnl{Planet.~Space~Sci.}}
\def\procspie{\jref@jnl{Proc.~SPIE}}

\def\ion#1#2{#1$\;${\small\rm\@Roman{#2}}\relax}

\newcount\lecurrentfam
\def\LaTeX{\lecurrentfam=\the\fam \leavevmode L\raise.42ex
\hbox{$\fam\lecurrentfam\scriptstyle\kern-.3em A$}\kern-.15em\TeX}
\def\sizrpt{
(\fontname\the\font): em=\the\fontdimen6\font, ex=\the\fontdimen5\font
\typeout{
(\fontname\the\font): em=\the\fontdimen6\font, ex=\the\fontdimen5\font
}}
\definecolor{dark}{rgb}{ 0.34, 0.09, 0.49}

\titlespacing*{\section}{0pt}{*3}{*0.7}
\usepackage{fancyhdr,url}
\usepackage{transparent}
\usepackage{tablefootnote}
\usepackage{mathpazo}

\definecolor{MyRed}{rgb}{0.9,0.0,0.0} 
\definecolor{MyDarkRed}{rgb}{0.6,0.0,0.0} 
\definecolor{MyLightRed}{rgb}{1.0,0.8,0.8} 
\definecolor{MyLightPink}{rgb}{1.0,0.9,0.9} 
\definecolor{MyLightOrange}{rgb}{1.0,0.8,0.5} 
\definecolor{MyPink}{rgb}{1.0,0.08,0.45} 
\definecolor{MyDarkBlue}{rgb}{0,0.08,0.45} 
\definecolor{MyDarkGreen}{rgb}{0,0.5,0.0} 
\definecolor{MyLightBlue}{rgb}{0.97,0.97,1.0} 
\definecolor{MyLightGreen}{rgb}{0.8,1.0,0.8} 
\definecolor{MyWarmYellow}{rgb}{1.0,0.95,0.85} 
\definecolor{MyMediumBlue}{rgb}{0.7,0.72,1.0} 

\newcommand{\edt}[1]{#1}

\begin{document}

\noindent
\begin{raggedright}
{\Huge\bfseries 
The energetic atmosphere of the Sun: \\[0.5em]
\Large How do small-scale dynamic processes heat and structure \\the solar atmosphere?
}\end{raggedright}\\

\noindent
Sven Wedemeyer$^1$, Mats Kirkaune$^1$, Miroslav Barta$^2$,
Astrid M. Veronig$^3$, Stanislav Gun\'ar$^2$, Roman Braj\v{s}a$^4$,
Manuela Temmer$^5$, Valery Nakariakov$^6$, Stephen White$^7$,
Bart De Pontieu$^8$, Maryam Saberi$^1$, Atul Mohan$^9$,
Masumi Shimojo$^{9}$, Davor Sudar$^4$, Galina Motorina$^{10}$,
Maria Loukitcheva$^{11}$, Paulo Simões$^{12,13}$, 
Stefaan Poedts$^{14,15}$, 
Raghavendra Sahai$^{16}$

\vspace{1em}
{
\footnotesize{
\noindent
$^1$ Rosseland Centre for Solar Physics, Institute of Theoretical Astrophysics, University of Oslo, Norway \\
$^2$ Astronomical Institute, The Czech Academy of Sciences, Ond\v{r}ejov, Czechia \\
$^3$ University of Graz, Institute of Physics, Kanzelh{\"o}he Observatory for Solar and Environmental Research, Graz,\,Austria \\
$^4$ University of Zagreb, Faculty of Geodesy, Hvar Observatory, Croatia \\
$^5$ Institute of Physics, University of Graz, Austria \\
$^6$ Centre for Fusion, Space and Astrophysics, Department of Physics, University of Warwick, UK \\
$^7$ Space Vehicles Directorate, Air Force Research Laboratory, Kirtland AFB, NM, USA; Department of Physics and Astronomy, University of New Mexico, Albuquerque, USA \\
$^8$ Lockheed Martin Solar \& Astrophysics Laboratory, Palo Alto, USA \\
$^9$ ALMA project, National Astronomical Observatory of Japan, National Institute of Nature Science, Tokyo, Japan \\
$^{10}$ Central Astronomical Observatory at Pulkovo of Russian Academy of Sciences; Ioffe Institute, Polytekhnicheskaya, St. Petersburg, Russia \\
$^{11}$ Max Planck Institute for Solar System Research, Goettingen, Germany; Special Astrophysical Observatory, Nizhnii Arkhyz, Russia \\
$^{12}$ Center for Radio Astronomy and Astrophysics Mackenzie, Mackenzie Presbyterian University, São Paulo, Brazil \\
$^{13}$ UPA School of Physics and Astronomy, University of Glasgow, UK \\
$^{14}$ CmPA, Dept.\ of Mathematics, KU Leuven, Celestijnenlaan 200B, 3001 Leuven, Belgium\\
$^{15}$ Institute of Physics, University of Maria Curie-Skłodowska, Pl.\ Marii Curie-Skłodowskiej 1, 20-031 Lublin, Poland \\
$^{16}$ Jet Propulsion Laboratory, 4800 Oak Grove Drive, California Institute of Technology, Pasadena, CA 91109, USA
}}

\begin{tcolorbox}[boxsep=3pt,left=0pt,right=0pt,top=1pt,bottom=1pt,colframe=boxcolor,colback=boxcolor]
\justify
\textbf{Abstract:} 
Understanding the mechanisms that heat the Sun's chromosphere and corona to temperatures far exceeding those of the solar photosphere below remains one of the central challenges in solar physics. Small-scale, rapidly evolving processes -- including 
shocks, turbulence, magnetohydrodynamic waves,
reconnection-driven jets, and nano-flares -- are thought to contribute significantly to the heating budget. 
However, direct observational constraints on their spatial, temporal, and \edt{spectral} signatures remain hard to obtain. \edt{At millimeter wavelengths, optically thick thermal emission provides a comparatively direct measure of gas temperature.} While the Atacama Large Millimeter/submillimeter Array (ALMA) already provides unique diagnostic capabilities, offering temperature diagnostics and high-cadence imaging, the current capabilities are still limiting the study of such dynamic, small-scale phenomena.  
\textbf{Improved spatial resolution, enhanced uv-coverage, and simultaneous multi-band observations, as proposed for ALMA2040, would enable transformative progress} in capturing dynamic processes across multiple heights and \edt{enable time-dependent three-dimensional thermal and magnetic tomography of the chromosphere. Combined with complementary diagnostics and simulations, these measurements would test how small-scale processes contribute to atmospheric heating.}
\end{tcolorbox}

\newpage{}

\begin{figure} 
\vspace{-2mm}
\centering
\includegraphics[width=\textwidth]{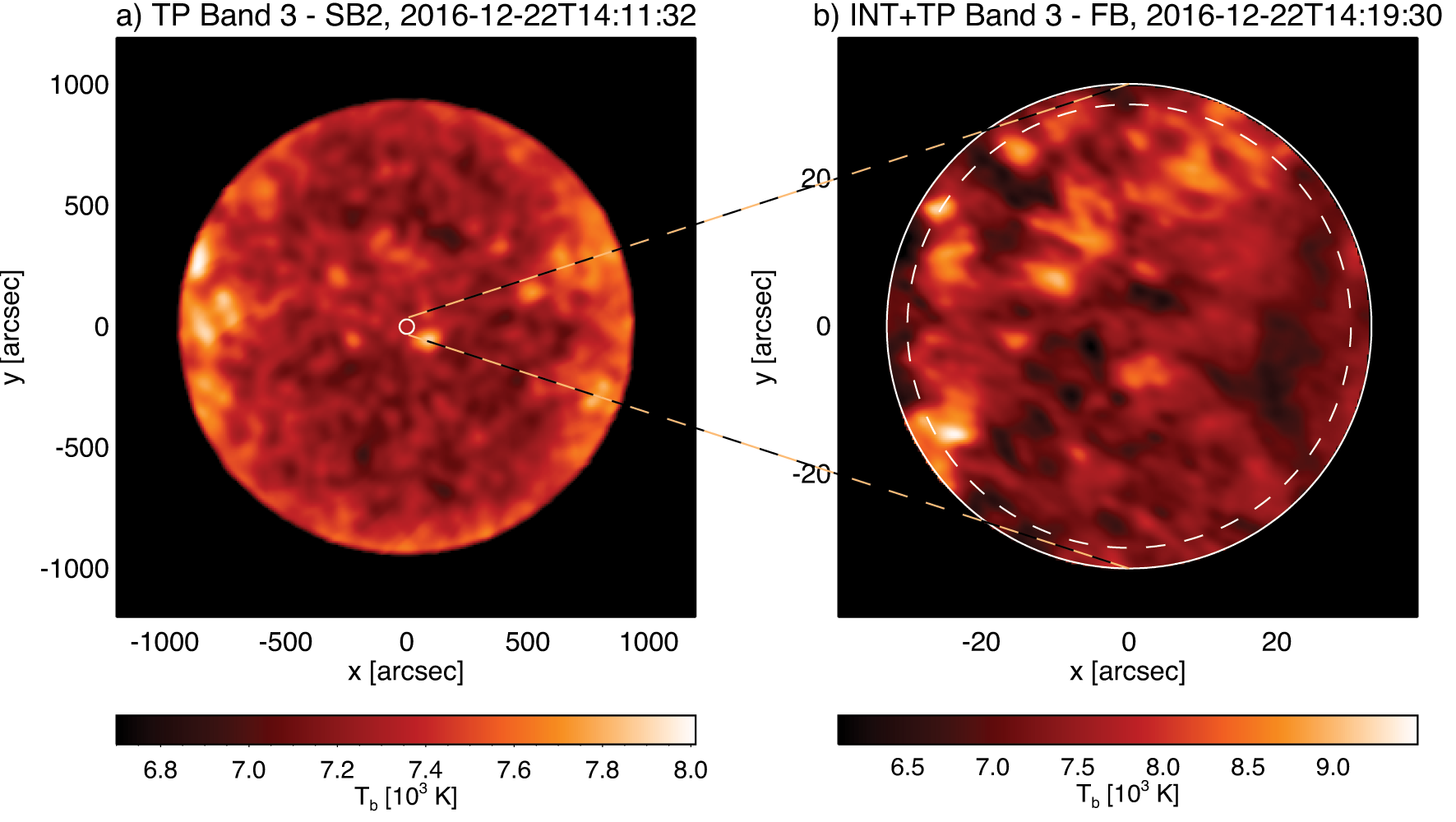}
\vspace*{-6mm}
\caption{\it Examples of Band 3 observations of the Sun including a single-dish TP map (left) and an interferometric map (right). From \citet{2020IAUS..354...24W}. Data set: ADS/JAO.ALMA \#2016.1.00423.S}
\vspace{-1mm}
\end{figure}

\section{Scientific Context and Open Questions}
Understanding the physical processes responsible for heating the outer layers of the solar atmosphere remains one of the crucial open questions in solar physics since the initial discovery in the 1940. 
The corona\footnote{The solar atmosphere consists of the photosphere, chromosphere, transition region, and corona, although it is best viewed as a coupled and highly dynamic system \citep{2009SSRv..144..317W}.} exhibits temperatures that exceed those of the photosphere by several orders of magnitude -- a phenomenon that is also observed in other stars \citep{Schrijver1987}. 
Importantly, the energy sustaining the high coronal temperatures is transported from the solar interior and photosphere through the chromosphere, a layer that remains notoriously difficult to observe.
\edt{Its rapid evolution and complex non-LTE diagnostics hinder inference of temperatures, densities and magnetic fields. Limited resolution also obscures magnetic field gradients needed to estimate currents and the connection between large-scale energy storage and small-scale dissipation. 
While some of these limitations also affect ALMA, further upgrades, as proposed for ALMA2040, promise substantial improvements of its diagnostic capabilities.} 

The Sun displays active regions that are dominated by strong magnetic fields and produce flares. The energetic particles and intense radiation released by flares only heat the atmosphere locally and briefly. In contrast, the quiet Sun hosts more frequent micro- and nano-flares, likely driven by small-scale magnetic reconnection in weaker field regions, which may contribute to a more continuous background heating.
The current assumption is that continuous heating can be attributed to ubiquitous, highly efficient energy-transport and dissipation mechanisms operating on very small spatial and temporal scales. 
Magnetic fields, convective motions, and plasma instabilities generate a wide range of dynamic phenomena, such as various magnetohydrodynamic wave modes, shocks, turbulent motions, and reconnection-driven events 
\citep{1998SoPh..182..349B,
2006ApJ...648L.151J, 
2006ApJ...647L..73H,
2007Sci...318.1574D, 
2008A&A...479..213B,
2008ApJ...681L.125S,
2008ApJ...680.1542H,
2008ApJ...676L..73V,
2008A&A...480..839F,
2009Sci...323.1582J,
2010ApJ...724.1542K,
2010ApJ...711..164V,
2012ApJ...753..161M,
2012Natur.486..505W,
2014ApJ...796L..23S,
2014Sci...346D.315D,
2020ARA&A..58..441N}. 
Unfortunately, present observational capabilities remain insufficient for a conclusive assessment of their efficiency and quantitative contributions.
\edt{In this context, the scientific potential of improved diagnostic capabilities is clearly demonstrated by recent DKIST observations, the first 4-m class solar telescope at optical wavelengths: The increased spatial resolution resulted in the discovery of ubiquitous photospheric Kelvin--Helmholtz instabilities at magnetic-flux boundaries \citep{Kuridze2026}. 
\citet{Nykyri2026} estimates that the associated shear-energy reservoir could contribute to coronal heating if sufficient energy is converted into magnetic free energy and transported upward. These results identify a promising mechanism like those mentioned above. The remaining questions central to the coronal heating problem, however, concern the transmission through the chromosphere, dissipation at coronal heights and overall contribution to the heating budget.} 

\begin{figure} 
\vspace{-3mm}
\floatbox[{\capbeside\thisfloatsetup{capbesideposition={right,top},capbesidewidth=5cm}}]{figure}[\FBwidth]
{\caption{\it Composite full-disk image of the Sun revealing the intricate atmospheric structure and the extension of the corona. The inset qualitatively illustrates the substantial change in mass density and temperature with height and the related atmospheric heating. Image: SOHO/NASA, adapted by S.\,Wedemeyer (2012). }
}
{\includegraphics[width=11cm]{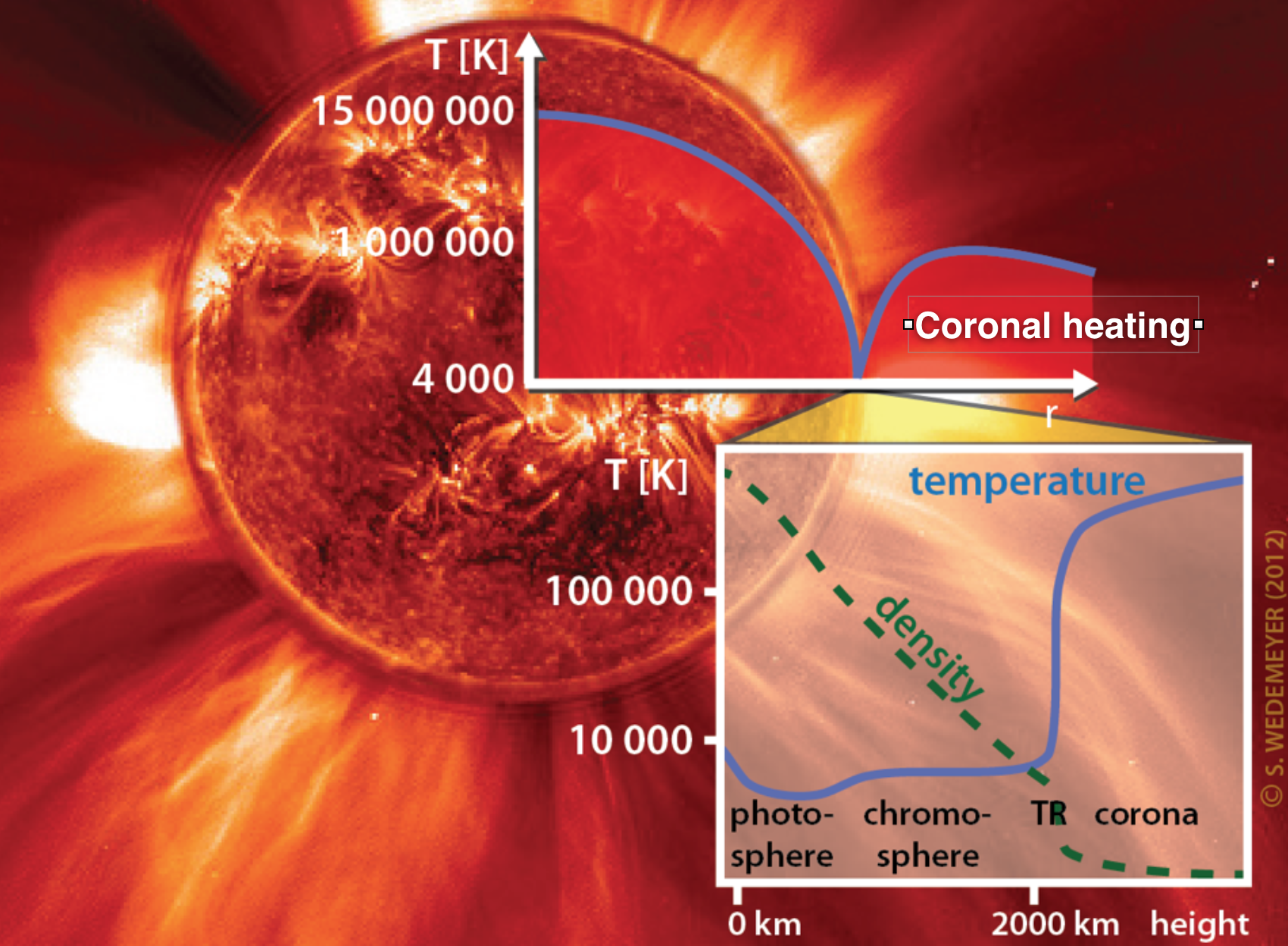}}
\end{figure}

Millimeter and submillimeter observations of the solar atmosphere (Fig.\,1) offer a crucial diagnostic advantage. Unlike many traditional chromospheric diagnostics in the UV, whose interpretation is complicated by non-LTE effects and intricate spectral line formation,
\edt{the millimeter continuum provides a comparatively direct temperature diagnostic where optically thick thermal emission dominates} 
\citep{2016SSRv..200....1W,2017A&A...605A.125S}.
In addition, the circular-polarisation spectrum can constrain the line-of-sight magnetic-field component (under appropriate magneto-ionic and radiative transfer assumptions),  
which is fundamental for reconstructing the three-dimensional atmospheric structure  \citep{2017A&A...601A..43L}. 
Nevertheless, limitations in spatial resolution, instantaneous $uv$-coverage, and the absence of simultaneous multi-band observations currently hinder the full exploitation of these capabilities. Consequently, small-scale events evolve rapidly and frequently occur below the current spatial resolution limits of even the most advanced observatories at (sub)mm wavelengths. As a result, the relative contributions of wave dissipation, magnetic reconnection, and turbulent cascade processes to the total energy budget remain uncertain. It is also unclear how these mechanisms interact across atmospheric layers, how energy is transported upward, and how magnetic topology shapes the frequency and efficiency of heating events.

\section{Science Case for the 2040s}
\label{sec:science}

The central scientific question for the coming decades concerns the exact way in which small-scale dynamic processes deposit energy into the chromosphere and how this energy subsequently contributes to \textbf{atmospheric heating} (Fig.~2). 
Addressing this challenge requires observations that resolve thermal structures with high spatial fidelity, trace their rapid temporal evolution, and provide sensitivity to vertical atmospheric stratification.
Achieving an angular resolution of 0.1'' or better is essential, as \edt{0.1'' corresponds to approximately 70\,km on the Sun, resolving finer structure in chromospheric fibrils and flux tubes and constraining the observable consequences of reconnection, even where dissipative scales remain unresolved.} 


An essential capability would be to observe simultaneously across a large frequency range or at least in multiple frequency bands.
Such capability is indispensable for capturing the height dependence of propagating disturbances and for \textbf{reconstructing the dynamically evolving 3D thermal and magnetic structure of the solar atmosphere}.
In combination with rapid temporal cadence, full-polarisation multi-band observations would allow the tracking of shocks, wave fronts, reconnection signatures, and thermal perturbations as they propagate through the atmosphere.
This would enable \edt{model-constrained estimates} of vertical energy fluxes and permit evaluation of the relative importance of different heating mechanisms.
These advances would place ALMA at the centre of coordinated multi-instrument campaigns involving observatories including \edt{DKIST \citep{Rimmele2020}, the European Solar Telescope (EST; \citealt{QuinteroNoda2022}), and the space missions MUSE \citep{DePontieu2020MUSE} and Solar-C \citep{Shimizu2019}}.
Understanding the heating of the solar atmosphere has implications for the atmospheric heating and activity of other stars and as such for the environments that host exoplanets.

\textbf{ALMA's distinctive contribution} to this multi-instrument approach lies in its comparatively direct \textbf{measurement of chromospheric temperatures} through (sub)-mm continuum emission.
\edt{For optically thick thermal emission, the LTE continuum source function makes brightness temperature a contribution-weighted diagnostic of electron (and thus gas)  temperature. However, the continuum opacity and formation height depend on electron density and non-LTE, time-dependent hydrogen ionisation \citep{Loukitcheva2015}. Joint multiwavelength inversions should therefore be validated against non-equilibrium radiation-MHD simulations, with uncertainties in temperature, density and height propagated into energy estimates.}
ALMA thus provides independent constraints on the thermal structure of chromospheric layers that are difficult to diagnose unambiguously with other instruments.
Together with complementary optical, UV, and EUV diagnostics, these measurements would provide a more complete picture of the physical conditions governing atmospheric heating and dynamics.
\edt{To connect local events to coronal heating, these observations would be combined with ensembles of 3D radiation-MHD simulations spanning magnetic environments and heating mechanisms. Forward modelling through realistic instrumental responses would test predicted temperature changes, time lags and spatial correlations against simultaneous ALMA, optical, UV and EUV measurements. Spectroscopic densities and velocities and magnetic field constraints would help quantify energy fluxes and their partition between local heating and upward transport. The energy balance must include ionisation, radiative losses, conduction and mass flows. 
Combining inferred energy fluxes with event energies and occurrence rates, while accounting for unresolved structure, would connect small-scale activity to large-scale energy budgets and test whether the energy reaching the corona offsets its radiative and conductive losses.}

\clearpage
In combination with complementary co-observations and data-constrained modelling, ALMA2040 would enable or substantially improve constraints on the following: 
\begin{itemize} 
\setlength{\itemsep}{1pt}
\item Energy flux carried by propagating MHD waves.
\item Frequency distribution of impulsive heating events.
\item Vertical transport of thermal energy through the chromosphere.
\item Magnetic topology associated with heating sites.
\item Contribution of wave dissipation versus reconnection to atmospheric heating.
\end{itemize}
 
These investigations would also advance understanding of the solar origins of space weather. Chromospheric heating and dynamics are closely coupled to coronal processes driving solar wind acceleration and magnetic energy release. By constraining the thermal structure and evolution of selected regions, ALMA2040 would clarify the physical conditions underlying solar wind generation, flare triggering, and solar eruptions. Its limited field of view and observing constraints, however, make it a research instrument for investigating these processes, not a facility for operational space weather monitoring or forecasting.

\section{Technical Requirements}
\begin{tcolorbox}[boxsep=3pt,left=0pt,right=0pt,top=1pt,bottom=1pt,colframe=boxcolor,colback=boxcolor]
\noindent{\textbf{High-fidelity snapshot imaging with a cadence of \mbox{\small{$\leq$}0.1\,s} and an angular resolution of \mbox{{\small$\lesssim$}0''.1} (at 230\,GHz), combined with circular polarimetry and simultaneous multi-band coverage 
%
across an extended frequency range (at least 30--700\,GHz) is essential for resolving the relevant spatial and temporal scales, disentangling the physical processes behind the transport and dissipation of energy, and thus contributing to the solution of the coronal heating problem.
}}
\end{tcolorbox}

\begin{table}
\centering
\footnotesize
\setlength{\tabcolsep}{1pt}
\renewcommand{\arraystretch}{1.05}
\caption{\it Comparison of current ALMA and the proposed ALMA2040 capabilities.}
\label{tab:alma2040_capabilities}
\begin{tabular}{@{}p{0.48\linewidth}cc@{}}
\toprule
\textbf{Capability} &
\textbf{\shortstack{Current\\ALMA}} &
\textbf{ALMA2040} \\
\midrule
Resolution~at~$\nu =230$\,GHz& $\sim$0.7$''$ & 0.1$''$ \\
Simultaneous bands          & No              & Yes      \\
Full-Stokes                 & Limited         & Yes      \\
Snapshot fidelity           & Moderate        & High     \\
\mbox{Tomography}    & No              & Routine  \\
\bottomrule
\end{tabular}
\end{table}

Realising these goals requires substantial advances in observational capabilities, namely an interferometric array for (sub)mm wavelengths like \textbf{ALMA that enables aperture synthesis imaging of the Sun at longer baselines} and thus higher angular resolution than currently possible.
This also requires refined calibration strategies to ensure stable performance at longer baselines and enhanced instantaneous $uv$-coverage, accomplished by a substantially larger number of antennae and thus baselines.
As the rapid evolution of the solar atmosphere limits observations to snapshot imaging, improving the instantaneous $uv$-coverage is crucial for achieving high fidelity of snapshot imaging and thus the scientific potential of the resulting data.

There are two main limitations to achieving truly high-resolution solar observations: the rapid phase variations during scans and the very sparse sampling of visibilities in the $uv$-plane. While advanced self-calibration with sub-second integration\footnote{ESO ALMA Development Study \textit{``High-cadence imaging of the Sun''}, 2023, ALMA Memo 628} \citep{almamemo628} can partially address rapid phase variations, large phase errors still require pre-correction.
The sparse $uv$-coverage, however, remains a fundamental constraint\footnote{ESO ALMA Development Study \textit{``High resolution solar ALMA imaging''}} and can only be overcome by adding more antennas (see Sect.~\ref{sec:almasim}).
Accurate absolute brightness temperatures additionally require combination with single-dish data, yet there are challenges in combining current ALMA Total Power antennas \citep{2017SoPh..292...88W}  with interferometric observations.
An adequate facility building on ALMA would thus include water-vapour radiometers for phase pre-correction usable for solar observations, a larger ACA to increase $uv$-density, and a bigger multi-feed single-dish for recovering the missing large-scale flux.
The latter requirement could be met by co-observations with \mbox{AtLAST} \citep{2025A&A...694A.142M}
\edt{as the anticipated broad spectral and spatial coverage would complement the higher angular resolution of interferometric ALMA observations.}


The most critical advancement, however, is the implementation of \textbf{\underline{simultaneous} multi-band observations} across an extended frequency range (at least 30--700\,GHz).
Recording several bands simultaneously is essential for height discrimination and synchronised measurements across atmospheric layers.
The wavelength dependence of the (sub-)mm continuum opacity causes different frequencies to sample partially overlapping height ranges, providing a tomographic capability.
Simultaneity avoids ambiguities from temporal evolution between bands, while radiative-transfer-based inversions can combine complementary brightness temperature constraints to reconstruct the evolving vertical thermal structure.

Together with spatially resolved imaging, this would enable time-dependent three-dimensional thermal \edt{tomography of the solar chromosphere. 
Rather than a unique reconstruction at fixed geometric heights, the contribution functions at different frequencies overlap but also change rapidly in response to the chromosphere's dynamics. 
While the variations in principle limit the height resolution and introduce degeneracies between temperature, density and height, the  use of modern data inversion techniques, as already demonstrated for ALMA by  \citet{2018A&A...620A.124D},  and further developments based on physics-informed neural networks \citep{Diaz2025} have the potential of establishing solar tomography with ALMA as a powerful tool for studying the heating and dynamics as described in Sect.~\ref{sec:science}.}

Equally important is \textbf{full-Stokes polarimetry}, complementing thermal diagnostics with constraints on the chromospheric magnetic field.
Circular polarisation of the (sub-)mm continuum can constrain the line-of-sight magnetic field, while additional polarisation information, where measurable and physically interpretable, may constrain magnetic field geometry.
Combined with multi-band observations, \edt{simultaneous} magnetic-field measurements from other facilities, \edt{and the aforementioned data inversion techniques,} these capabilities would enable investigation of the relationship between magnetic structure, plasma dynamics, and energy release, helping establish how magnetic energy is transported, dissipated, and converted into atmospheric heating, dynamics, and activity.

\begin{figure} 
\vspace{-5mm}
{\caption{\it Simulated observations of the Sun with ALMA in Band~6 using different array configurations: a)~Current ALMA config.~3, and for ALMA2040 b)~H1, c)~H2, and d)~H3. Panel~e shows the corresponding brightness temperature distributions. The dynamic range of the images (triangles, dotted line) and the size of the synthesised beam (circles, dashed line) of the considered cases are shown as a function of maximum baseline length in panel~f.  }
}
\label{fig.sasim}
{\includegraphics[width=\textwidth]{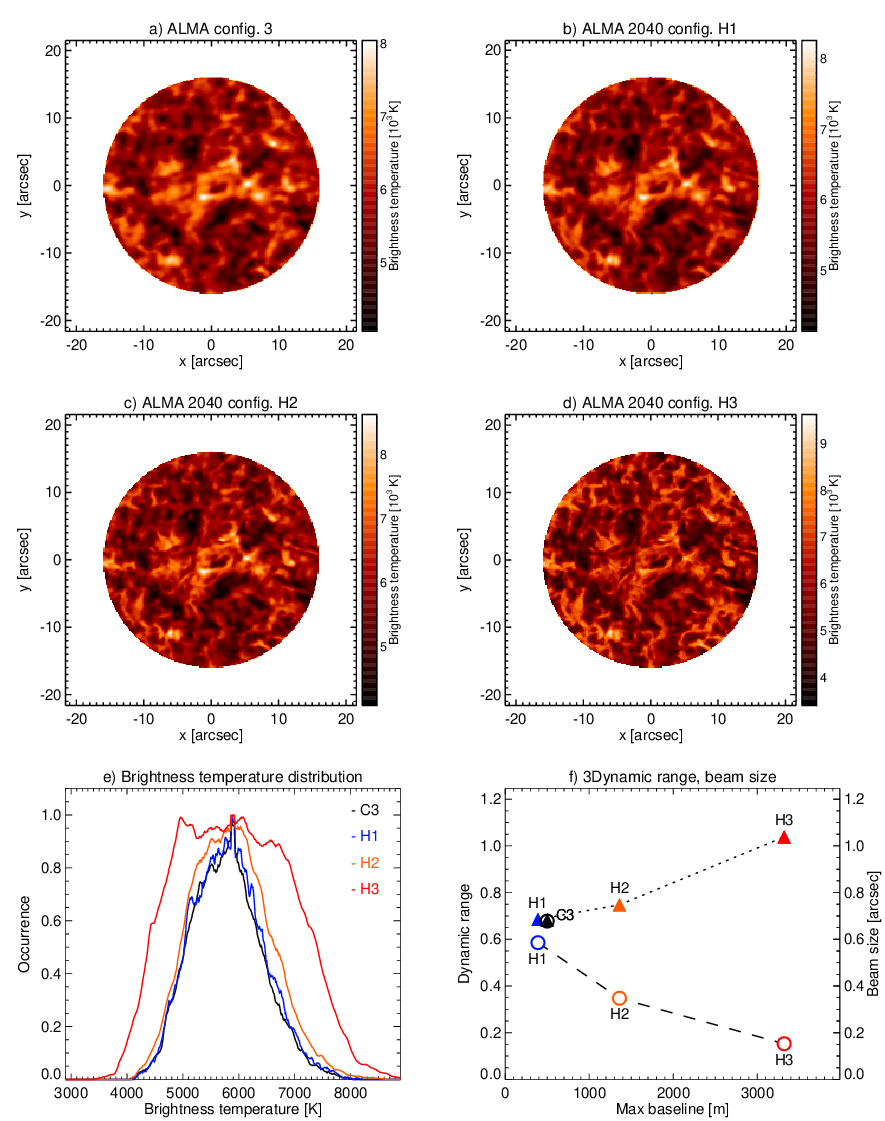}}
\end{figure}

\section{Telescope simulations}
\label{sec:almasim}
The gains in spatial resolution, imaging quality, and scientific accuracy are illustrated using the Solar ALMA Simulator (SASim), developed within the ESO-funded ALMA development study \textit{High-cadence Imaging of the Sun} \citep{almamemo628}. SASim uses synthetic millimetre brightness-temperature maps calculated with ART \citep{2021zndo...4604825D} from a Bifrost simulation \citep{2016A&A...585A...4C}. Based on the CASA \texttt{simobserve} task, these maps are converted into synthetic measurement sets that mimic solar ALMA observations, accounting for the adopted antenna configuration and prescribed atmospheric conditions. The resulting data are processed with the Solar ALMA Pipeline \citep[SoAP, ][]{2022A&A...659A..31H,2020A&A...635A..71W}, using the same deconvolution and imaging procedures as for real observations. Comparing the reconstructed images with the known input maps thus enables a direct assessment of imaging performance.

We compare the currently used C3 configuration with the proposed H1, H2, and H3 configurations. While ALMA comprises 50 12-m antennas in the main array, in addition to the ACA, each proposed configuration would contain 255 antennas. The maximum baselines are 390\,m, 1360\,m, and 3320\,m for H1, H2, and H3, respectively, compared with 500\,m for C3. Since imaging performance depends on the full baseline distribution and corresponding $uv$-coverage rather than the maximum baseline alone, this comparison is intended primarily as a first qualitative assessment.

The improvement in spatial resolution with H2 and H3 is immediately apparent in Fig.~3, with the effective resolution improving from 0.68'' for C3 to 0.15'' for H3, approaching that currently achieved by optical solar telescopes. The brightness-temperature histograms (Fig.~3e) likewise show that H3 recovers a broader range of temperatures, including higher and lower values increasingly suppressed by spatial smearing in the lower-resolution configurations.

As demonstrated by current ALMA observations, wider array configurations can provide higher angular resolution, but increasingly sparse $uv$-coverage eventually limits reliable snapshot imaging. The substantially larger number of antennas envisioned for ALMA2040 would mitigate this limitation, enabling robust imaging with more extended configurations. Our simulations show that, as the synthesised beam becomes smaller, the dynamic range continues to increase up to H3 (Fig.~3f). This suggests that reliable solar imaging could remain feasible with baselines exceeding 3\,km, owing to the much denser $uv$-coverage. Together with simultaneous multi-band observations and polarimetric capabilities, this would substantially advance ALMA's diagnostic potential for solar observations and enable the science cases outlined in Sect.~\ref{sec:science}.

\section{Conclusions}
Future enhancements of ALMA in angular resolution, instantaneous $uv$-coverage, and especially in simultaneous multi-band full-polarisation capability are vital for resolving the small-scale dynamic processes responsible for chromospheric and coronal heating. \edt{Combined with optical, UV and EUV diagnostics and advanced 3D simulations, these measurements would determine where energy is deposited, how much is transported upward, and which mechanisms can sustain the required heating. ALMA's distinctive thermal and magnetic constraints would thereby connect the evolution of small-scale structures to quantitative atmospheric energy budgets, thereby positioning}  ALMA at the forefront of efforts to address one of the most persistent open questions in solar physics: the coronal heating problem.

\section*{Acknowledgements}
This work was developed as part of the broader ALMA2040 community effort. 
We are grateful to Jaime de la Cruz Rodriguez and Tony Mroczkowski for their valuable feedback. 
We acknowledge the Lorentz Center and NOVA for supporting the community workshops and coordination activities that helped initiate and shape the ALMA2040 White Paper process.
SW acknowledges support by the Research Council of Norway through its Centres of Excellence scheme, project number 262622. 
MB, RB, and DS acknowledge support from the Croatian Science Foundation project IP-2025-02-3228 (Solar Chromosphere in ALMA's Eyes: Newly Ascending High-Resolution Interferometry and MHD Modeling to Reveal Hidden Wave Dynamics, SCi-AEneAs). 
S.P. is funded by the European Union (ERC-AdG agreement No 101141362, Open SESAME) and 
by the projects C16/24/010 (C1 project Internal Funds KU Leuven), G0B5823N and G002523N (WEAVE) (FWO-Vlaanderen), and 4000145223 (SIDC Data Exploitation (SIDEX2), ESA Prodex).
R.S.’s contribution 
was carried out at the Jet Propulsion Laboratory, California Institute of Technology, under a contract with NASA (80NM0018D0004).\\

\bibliographystyle{apj_mod}

\end{document}